\documentclass[pre,twocolumn, superscriptaddress,floatfix,amsmath,a4paper]{revtex4-2}
\usepackage{graphicx}
\usepackage[english]{babel}
\usepackage{amsmath,amssymb}
\usepackage[utf8]{inputenc}
\usepackage{psfrag}
\usepackage{bm}
\usepackage{xcolor}
\usepackage{hyperref}
\usepackage{appendix}
\usepackage[normalem]{ulem}

\begin{document}

\title{Fixation probabilities for multi-allele Moran dynamics with weak selection}

\author{Ian Braga}
\affiliation{Center for Advanced Systems Understanding (CASUS) -- Helmholtz-Zentrum Dresden-Rossendorf (HZDR), Untermarkt 20, Görlitz 02826, Germany}
\affiliation{Universidade Federal de Minas Gerais, Brazil}

\author{Lucas Wardil}
\affiliation{Universidade Federal de Minas Gerais, Brazil}

\author{Ricardo Martinez-Garcia}
\affiliation{Center for Advanced Systems Understanding (CASUS) -- Helmholtz-Zentrum Dresden-Rossendorf (HZDR), Untermarkt 20, Görlitz 02826, Germany}
\affiliation{ICTP South American Institute for Fundamental Research \& Instituto de F\'isica Te\'orica, Universidade
Estadual Paulista - UNESP, Brazil.}
\affiliation{Department of Ecology, Institute of Biosciences, University of São Paulo, São Paulo, Brazil }
 \email{r.martinez-garcia@hzdr.de}

\date{\today}

\begin{abstract}
Fixation probabilities are essential for characterizing stochastic evolutionary dynamics, but analytical results remain limited mainly to systems with two competing types. We develop a perturbative framework to compute fixation probabilities in multi-allele Moran processes under weak selection. Exploiting the general structure of the backward Fokker–Planck operator in this regime, we show that fixation probabilities admit a systematic expansion around their neutral solution. We first introduce the framework for $M$ competing alleles with multivariate polynomial fitness functions. We then apply it to three biologically motivated examples: a constant-fitness model, which we solve for an arbitrary number of alleles; a three-allele coordination game in which allele fitness increases with its frequency in the population; and a three-allele model of clonal interference between mutualistic alleles. These results extend the analytical understanding of fixation probabilities beyond pairwise interactions, establishing a framework for investigating multi-strategy stochastic evolutionary dynamics.
\end{abstract}

\maketitle

\section{Introduction}

Evolution is a stochastic process driven by the probabilistic extinction and fixation of traits. Consequently, deterministic models, such as the replicator equation \cite{TaylorJonker1978}, provide only a limited description of evolutionary dynamics \cite{Traulsen2009, Constable2016}. Even in the simple case of two neutral alleles competing within a population, deterministic and stochastic descriptions lead to contradictory results. Deterministic models predict indefinite coexistence, whereas stochastic approaches predict eventual dominance of one allele and the extinction of the other \cite{Nowak2006}. This fluctuation-driven dominance underlies neutral theory \cite{Kimura1983}, which explains seemingly counterintuitive phenomena, such as the atrophy of functional organs (e.g, blindness in cave fish \cite{Gore2018}), or the loss of flight in island parrots \cite{Wright2016}.

A minimal framework for studying such stochastic effects in finite populations is the Moran process, which describes how allele frequencies change in a population due to the joint influence of random genetic drift and selection arising from fitness differences \cite{Moran1958, Diaz2021}. In this model, the long-term fate of competing alleles is determined by whether a trait eventually goes extinct or takes over the population, which can be measured through fixation probabilities and fixation times \cite{McCandlish2015}.

For simple cases with only two competing alleles, closed-form expressions for fixation probabilities are well known under general fitness functions \cite{Ewens1963,KarlinTaylor1975} and for a wide range of evolutionary scenarios both in well-mixed and structured populations \cite{Allen2017,HindersinTraulsen2015, Lieberman2005, Ohtsuki2006}. These include strong-selection limits \cite{AltrockTraulsen2009}, fluctuating environments \cite{Saakian2019}, and frequency-dependent fitness functions arising in evolutionary game theory \cite{Nowak2006,Traulsen2009, AntalScheuring2006, McCandlish2015}. However, for more than two alleles, the dimension of the state space increases, and computing fixation probabilities becomes more challenging. For three competing alleles, for example, the state space is two-dimensional, and calculating fixation probabilities requires solving a boundary-value problem on a two-dimensional discrete lattice. More generally, for $M$ alleles, the boundary problem is posed in an $(M-1)$-dimensional discrete lattice.

The difficulty of obtaining analytical results in these high-dimensional problems has motivated the development of various approximate and perturbative approaches. These approaches have yielded bounds and inequalities on fixation probabilities
\cite{FerreiraNeves2018}
, as well as perturbative solutions for the related Wright–Fisher model
\cite{LessardLahaie2009}. Nevertheless, general analytical expressions for fixation probabilities in the multi-allele Moran process remain lacking, even in the weak-selection regime.

A useful way to overcome the technical difficulties posed by the discrete problem is to use a continuum approximation, replacing the discrete $(M-1)$-dimensional lattice boundary-value problem with a continuous formulation in terms of a Fokker-Planck equation on the $(M-1)$-dimensional simplex \cite{Gardiner2009, firstpaper, secondpaper}. This approximation maps fixation probabilities to the stationary solutions of the corresponding backward Fokker-Planck equation, with absorbing boundary conditions at the vertices of the simplex.

For two alleles and large population sizes, this continuum description reproduces the exact solutions of the discrete problem, providing a rigorous link between microscopic transition rates and the deterministic and stochastic components of the allele-frequency dynamics \cite{firstpaper}. However, for $M \ge 3$, the backward Fokker-Planck equation is a rarely analytically tractable high-dimensional partial differential equation defined in an $(M-1)$-dimensional space \cite{secondpaper, Gardiner2009}. Hence, generalizing the analytical results of the two-allele case to higher dimensions requires new approximations.

To fill this gap, we develop a perturbative approach to analytically compute fixation probabilities in weak-selection multi-allele Moran processes with frequency-dependent fitness functions. Our method relies on a systematic expansion of the fixation probability about its neutral solution and is valid, to first order in selection strength, when the fitness deviations are multivariate polynomial functions of the allele frequencies. To illustrate the generality of this approach, we apply it to three Moran processes with both constant and frequency-dependent fitness functions. For the latter, we study a three-species coordination game motivated by microbial cooperation and collective animal behavior \cite{Gore2009,Darch2012,Handegard2010}; and a model of mutualistic clonal interference inspired by interacting beneficial mutations in microbial evolution \cite{GerrishLenski1998,DesaiFisher2007}. In all cases, we validate the analytical predictions with Monte Carlo simulations, confirming the accuracy of our method across broad parameter ranges.

\section{Fixation probabilities in a multi-allele Moran process}

We consider a Moran process in a population of fixed size $N$. Each individual carries one of $M$ possible alleles $1, 2, \dots, M$ and $n_i$ is the number of individuals of type $i$, $\sum_{i=1}^M n_i = N$. Under these assumptions, the composition of the population is entirely described by the state vector $\mathbf{n} = (n_1, ..., n_{M-1})$ and the condition $n_{M}=N-\sum_{i=1}^{M-1} n_i$.

This state vector changes in time according to the following update rule. In a short time interval $dt$, one individual of type $i$ is chosen to reproduce with probability $F_i(\mathbf{n})$ and produces a sibling that is also of type $i$. To ensure that the total population size remains constant, a second individual is sampled with probability $D_j(\mathbf{n})$ and is replaced by the new individual. We define $f_i$ as the fitness of type $i$ in the current population state, understood broadly as a model-dependent measure of relative selective advantage and implemented here as a dimensionless reproductive weight. We assume that the reproduction probability is proportional to this weight, while all organisms have the same death probability,
\begin{equation}
F_i(\mathbf{n}) = \frac{f_i n_i}{\sum_k f_k n_k};\qquad D_j(\mathbf{n}) = \frac{n_j}{N}. \nonumber
\end{equation}

Assuming the birth and death events are independent, this update rule can be expressed in terms of replacement rates for the components of the state vector. We define these rates as $T_{i\rightarrow j}(\mathbf{n})=\lambda F_i(\mathbf{n})\, D_j(\mathbf{n})$, where $\lambda=1$ sets the characteristic time scale. The subscript $i\rightarrow j$ in the left side indicates a transition from a state $\mathbf{n}$ to a new state $\mathbf{n}'$ in which a copy of an individual of type $i$ replaces one individual of type $j$. The resulting state is thus $\mathbf{n}'=\mathbf{n}+\bm{\delta}_i-\bm{\delta}_j$, where $\bm{\delta}_k$ with $k=1,\ldots,M-1$ denotes the unit vector with components $(\bm{\delta}_k)_\ell = \delta_{k\ell}$.

Considering this updating rule, the probability that the system is in a state $\mathbf{n}$ at time $t$, $P(\mathbf{n},t)$, changes in time according to a Master equation
\begin{equation}\label{eq:master}
 \frac{\partial P(\mathbf{n},t)}{\partial t} = \sum_{\substack{i,j=1\\ i\neq j}}^{M} \left(E_{i}^{-}E_{j}^{+} - 1\right)\left[T_{i\rightarrow j}(\mathbf{n})P(\mathbf{n},t)\right],
\end{equation}
where we have introduced the step operators \cite{Toral2014},
\begin{equation}\label{eq:step}
E_{i}^{\pm}[f(\mathbf{n})] = f(\mathbf{n}\pm\bm{\delta}_i) \quad i=1,\ldots,M
\end{equation}
and $\bm{\delta}_M=0$ to simplify the notation while accounting for the fact that updates in the state vector $\mathbf{n}$ involving the $M$-th allele only change one component of $\mathbf{n}$.

In the limit where the population is sufficiently large, we can define continuous allele frequencies $x_i=n_i/N$ and perform a standard Kramers-Moyal expansion of the Master equation \eqref{eq:master} to obtain its corresponding Fokker-Planck diffusive approximation \cite{Gardiner2009} (see Appendix\,\ref{sec:appKramersMoyal} for details),
\begin{eqnarray}\label{eq:FPE}
 \frac{\partial \rho(\mathbf{x},t)}{\partial t} &=& -\sum_{k=1}^{M-1}\frac{\partial}{\partial x_k}\bigg(A_k(\mathbf{x})\rho(\mathbf{x},t)\bigg) \nonumber \\
 &&+ \frac{1}{2}\sum_{k,\ell=1}^{M-1}\frac{\partial^2}{\partial x_k \partial x_\ell} \bigg(B_{k\ell}(\mathbf{x})\rho(\mathbf{x},t)\bigg),
\end{eqnarray}
where the drift coefficients are given by
\begin{equation}\label{eq:Ak}
A_k(\mathbf{x}) = \frac{1}{N}
\sum_{i=1}^{M} \left[ T_{k\to i}(\mathbf{x}) - T_{i\to k}(\mathbf{x}) \right]
\end{equation}
and the elements of the diffusion matrix are
\begin{eqnarray}
B_{kk}&=&\frac{1}{N^2}
\sum_{\substack{i=1 \\ i\neq k}}^M\left[ T_{k\to i}(\mathbf{x})+ T_{i\to k}(\mathbf{x})\right], \label{eq:Bkk} \\
B_{k\ell}&=&-\frac{1}{N^2}\left[T_{k\to \ell}(\mathbf{x}) + T_{\ell\to k}(\mathbf{x})\right] \quad \text{for} \ k\neq \ell. \label{eq:Bkl}
\end{eqnarray}

After deriving the Fokker-Planck equation for the multi-allele Moran process, we can obtain each allele's fixation probability using the corresponding backward Fokker-Planck equation \cite{Gardiner2009}. At long times, the probability mass eventually flows towards one of the $M$ absorbing states, each corresponding to the fixation of one of the $M$ competing alleles. Mathematically, these states correspond to configurations of $\mathbf{x}$ where one component is equal to $1$ and all the others are equal to $0$ and $(x_1,\dots,x_{M-1})=(0,\dots,0)$, corresponding to fixation of allele $M$.

We are interested in the allele-specific fixation probabilities given an initial condition for the population composition $\mathbf{x}$,
\begin{equation}
\phi_i(\mathbf{x}) = \Pr(\text{fixation of allele } i \mid \mathbf{x}) \quad i = 1,\ldots,M. \nonumber
\end{equation}
Interpreting Eq.\,\eqref{eq:FPE} as a continuity equation for the probability density, fixation probabilities are given by the stationary backward Fokker-Planck equation \cite{Gardiner2009}
\begin{equation}\label{FixationEq}
\mathcal{L}\phi_i(\mathbf{x})=0,
\end{equation}
where $\mathcal{L}$ is a backward Fokker-Planck operator
\begin{equation}\label{eq:backward_operator}
\mathcal{L}
=
\sum_{k=1}^{M-1} A_k(\mathbf{x})\,\frac{\partial}{\partial x_k}
+\frac{1}{2}\sum_{k,\ell=1}^{M-1} B_{k\ell}(\mathbf{x})\,\frac{\partial^2}{\partial x_k\partial x_\ell},
\end{equation}
with boundary conditions set by the absorbing structure of the Moran process,
\begin{equation}\label{eq:fixation-boundary}
\phi_i(\mathbf{x}) =
\begin{cases}
1 & \text{if } \mathbf{x} = \bm{\delta}_i, \\
0 & \text{if } \mathbf{x} = \bm{\delta}_j,\; j \neq i.
\end{cases}
\end{equation}
Thus, obtaining the fixation probabilities for each allele is equivalent to solving a stationary multi-dimensional advection-diffusion equation in a $M-1$-dimensional simplex.

\section{Perturbative approach to fixation probabilities under weak selection} \label{sec:general-perturbative}

In the weak-selection regime, the drift coefficients in Eq.\,\eqref{FixationEq} are linear in the selection intensity, and the diffusion matrix reduces to its neutral form. More specifically, writing the fitness functions as
\begin{equation}\label{eq:weaksel_fitness_general}
f_i(\mathbf{x}) = 1 + s\,\pi_i(\mathbf{x}),\qquad s\,\pi_i(\mathbf{x})<\mathcal{O}(N^{-1}),
\end{equation}
and the mean contribution to fitness due to selection as
\begin{equation}\label{eq:barpi_def}
\bar\pi(\mathbf{x})=\sum_{j=1}^{M} x_j \pi_j(\mathbf{x}),
\end{equation}
the elements of the drift vector and the diffusion matrix are, respectively,
\begin{equation}\label{eq:Ak_weak_general}
A_k(\mathbf{x}) \simeq \frac{s}{N}\,x_k\big[\pi_k(\mathbf{x})-\bar\pi(\mathbf{x})\big],
\end{equation}
and
\begin{equation}\label{eq:B_neutral_general}
B_{k\ell}(\mathbf{x})\simeq\frac{2}{N^2}
\begin{cases}
x_k(1-x_k), & k=\ell,\\[4pt]
- x_k x_\ell, & k\neq \ell,
\end{cases}
\end{equation}
where $k,\ell=1,\ldots,M-1$ (see Appendix~\ref{sec:appWeakSel} for a full derivation).

This structure of the drift vector and the diffusion matrix in the weak-selection regime enables a systematic perturbative solution for the fixation probabilities in Eq.\,\eqref{FixationEq}. To obtain these solutions, we first decompose the Fokker-Planck backward operator
\begin{equation}\label{eq:L_decomp}
\mathcal{L} \simeq \mathcal{L}^{(0)} + \,\mathcal{L}^{(s)},
\end{equation}
where
\begin{equation}\label{eq:L0}
\mathcal{L}^{(0)}
=
\frac{1}{2}\sum_{k,\ell=1}^{M-1} B_{k\ell}(\mathbf{x})\,\frac{\partial^2}{\partial x_k\partial x_\ell},
\end{equation}
and
\begin{equation}\label{eq:L1}
\mathcal{L}^{(s)}
=
\sum_{k=1}^{M-1} A_k(\mathbf{x})\,\frac{\partial}{\partial x_k}.
\end{equation}

Based on the structure of the operator $\mathcal{L}$, we propose a similar decomposition for the first-order approximation of the fixation probabilities,
\begin{equation}\label{eq:phi_expansion}
\phi_i(\mathbf{x}) \simeq \phi_i^{(0)}(\mathbf{x}) + \,\phi_i^{(s)}(\mathbf{x}).
\end{equation}

Substituting \eqref{eq:L_decomp}--\eqref{eq:phi_expansion} into \eqref{FixationEq} and neglecting terms of order $\mathcal{O}(s^2)$ yields a recursive set of equations for each of the $s$-order contributions in which we have decomposed the fixation probability,
\begin{eqnarray}
\mathcal{L}^{(0)}\phi_i^{(0)}&=&0 \label{eq:order0} \\
\mathcal{L}^{(0)}\phi_i^{(s)} &=& -\mathcal{L}^{(s)}\phi_i^{(0)}.\label{eq:order1}
\end{eqnarray}

\noindent \textbf{Neutral solution and perturbative correction}. In the neutral Moran process, all alleles are equivalent, and fixation probabilities are given by their initial frequencies \cite{Nowak2006},
\begin{equation}\label{eq:neutral_solution}
\phi_i^{(0)}(\mathbf{x})=x_i,\qquad i=1,\ldots,M.
\end{equation}

\noindent These solutions satisfy Eq.\,\eqref{eq:order0} together with the absorbing boundary conditions (note that $\mathcal{L}^{(0)}x_i=0$ because $\mathcal{L}^{(0)}$ only contains second-order derivatives). At the same time, this neutral fixation probabilities provide the explicit form of the inhomogeneity in Eq.\,\eqref{eq:order1}
\begin{equation}\label{eq:forcing_general}
-\mathcal{L}^{(s)}\phi_i^{(0)}
=
-\sum_{k=1}^{M-1} A_k(\mathbf{x})\,\frac{\partial}{\partial x_k}x_i
=
- A_i(\mathbf{x}),
\end{equation}
for $i = 1, ..., M-1$. For $i = M$ the normalization of the frequency imposes $\phi_M=1-\sum_{i=1}^{M-1}\phi_i$.

Replacing these results in Eq.\,\eqref{eq:order1}, we obtain a linear elliptic problem for the first-order corrections to the fixation probabilities $\phi_i^{(s)}$
\begin{equation}\label{eq:PDE_phi1_general}
\frac{1}{2}\sum_{k,\ell=1}^{M-1} B_{k\ell}(\mathbf{x})\,\frac{\partial^2}{\partial x_k\partial x_\ell}\phi_i^{(s)}(\mathbf{x})
=
-\frac{s}{N}\,x_i\big[\pi_i(\mathbf{x})-\bar\pi(\mathbf{x})\big].
\end{equation}

Additionally, because the zeroth-order solution $\phi_i^{(0)}(\mathbf{x})=x_i$ already satisfies the absorbing boundary conditions in Eq.\,\eqref{eq:fixation-boundary}, the first-order perturbative correction must vanish at every absorbing vertex. Therefore, we impose $\phi_i^{(s)}(\mathbf{x}) = 0$ at all absorbing vertices. This condition ensures that the full solution
\begin{equation}
\phi_i(\mathbf{x}) = x_i + \mathcal{O}(s)
\end{equation}
remains consistent with certain fixation when $\mathbf{x}=\bm{\delta}_i$ and impossible fixation when the initial condition is such that $x_i = 0$.\\

\noindent \textbf{Multivariate polynomial ansatz and general solution}. Equation~\eqref{eq:PDE_phi1_general} makes the structure of the fixation probability in the weak-selection limit explicit. The backward neutral operator $\mathcal{L}^{(0)}$ acts on the unknown perturbative correction, $\phi_i^{(s)}$, while the right-hand side is an explicit polynomial forcing term generated by the selection-induced drift acting on the neutral solution $\phi_i^{(0)}=x_i$.

This structural property of Eq.\,\eqref{eq:PDE_phi1_general} allows us to propose an ansatz for the first-order perturbative contribution to the fixation probability. First, we assume that the fitness deviations on the right side, $\pi_i(\mathbf{x})$ are polynomial functions of the $M-1$ independent allele frequencies, with each variable appearing with degree at most $d$. In this case, they can be written as
\begin{equation}\label{eq:pi_poly_clean}
\pi_i(\mathbf{x})
=
\sum_{\alpha_1=0}^{d}
\cdots
\sum_{\alpha_{M-1}=0}^{d}
c_{i,\alpha_1,\ldots,\alpha_{M-1}}
\,x_1^{\alpha_1}\cdots x_{M-1}^{\alpha_{M-1}},
\end{equation}
Under this assumption, the forcing term
\begin{equation}
-\frac{s}{N}\,x_i\big[\pi_i(\mathbf{x})-\bar\pi(\mathbf{x})\big]
\end{equation}
is also a polynomial in the same variables. On the left side, because $\mathcal{L}^{(0)}$ only contains second-order derivatives with polynomial coefficients in $x_1,\ldots,x_{M-1}$, this operator maps polynomials in these variables into polynomials with different coefficients. Therefore, consistency of Eq.\,\eqref{eq:PDE_phi1_general} requires that $\phi_i^{(s)}$ must be within the same polynomial class.

Under these assumptions, the difference $\pi_i(\mathbf{x})-\bar\pi(\mathbf{x})$ is a polynomial in $x_1,\ldots,x_{M-1}$, with each variable appearing with power at most $d+1$. Moreover, the right-hand side of Eq.\,\eqref{eq:PDE_phi1_general} contains an overall factor $x_i$. Since the correction must vanish when the focal allele is absent, we construct the ansatz
\begin{equation}\label{eq:ansatz_general_clean}
\phi_i^{(s)}(\mathbf{x})
=
N x_i
\sum_{\beta_1=0}^{d+1}
\cdots
\sum_{\beta_{M-1}=0}^{d+1}
C_{i,\beta_1,\ldots,\beta_{M-1}}
\,x_1^{\beta_1}\cdots x_{M-1}^{\beta_{M-1}}.
\end{equation}
The factor $x_i$ ensures that $\phi_i^{(s)}=0$ whenever $x_i=0$, while the prefactor $N$ accounts for the population-size scaling of the solution. Specifically, the neutral backward operator is of order $\mathcal{O}(N^{-2})$, whereas the forcing term is of order $\mathcal{O}(s/N)$. Therefore, the perturbative correction scales as $\mathcal{O}(Ns)$, and the coefficients $C_{i,\beta_1,\ldots,\beta_{M-1}}$ are of first order in the selection strength.

Substituting the ansatz \eqref{eq:ansatz_general_clean} into Eq.\,\eqref{eq:PDE_phi1_general} yields a polynomial identity on the $(M-1)$-dimensional simplex. Matching equal monomials produces a finite linear system for the unknown coefficients, while the absorbing boundary conditions fix the remaining degrees of freedom. Hence, whenever the fitness functions are polynomial in the independent allele frequencies with bounded degree in each variable, the weak-selection fixation probability admits an explicit polynomial construction obtained by coefficient matching.

The size of this linear system can also be determined explicitly. As shown in Appendix~\ref{sec:appPolynomialScaling}, each exponent $\beta_k$ can take $d+2$ possible values, from $0$ to $d+1$. The ansatz therefore contains at most
\begin{equation}
K(M,d)=(d+2)^{M-1}
\end{equation}
unknown coefficients for each focal allele. Since only $M-1$ fixation probabilities are independent, the complete construction contains at most $(M-1)K(M,d)$ coefficients, separated into $M-1$ independent systems. Importantly, this number does not depend on the population size $N$. For low-degree fitness functions and moderate numbers of alleles, the method remains numerically accessible and provides the complete fixation-probability field.

\section{Applications to specific fitness functions}

After developing a general method to calculate fixation probabilities in multi-allele Moran processes under weak selection, we apply it to three biologically relevant variants. We first consider the simplest case with constant fitness but allele-specific selective advantage. Then, we study two frequency-dependent fitness scenarios: a coordination game, where fitness increases with allele frequency, and a model of clonal interference between mutualistic alleles.

\subsection{Constant fitness}
\label{subsec:const}

We start with the simplest case of allele-specific constant fitness $\pi_i(\mathbf{x})=p_i$ and $M = 3$. Defining $s_i\equiv s p_i$, the fitness functions are
\begin{equation}
f_1 = 1 + s_1,\qquad
f_2 = 1 + s_2,\qquad
f_3 = 1 + s_3,
\end{equation}
with $s_i <\mathcal{O}(N^{-1})$. After eliminating $x_3$ using the fixed population size condition, $x_3=1-x_1-x_2$, the dynamics is defined in the $(x_1,x_2)$ simplex.

Using the general weak-selection expressions derived in Eqs.~\eqref{eq:Ak_weak_general}--\eqref{eq:B_neutral_general}, the drift components are
\begin{equation}\label{eq:Ak_const}
A_i=\frac{x_i}{N}\big(s_i-\bar s\big),\qquad i=1,2,
\end{equation}
where
\begin{equation}\label{eq:sbar_const}
\bar s=\sum_{i=1}^{3}x_is_i,
\qquad
x_3=1-x_1-x_2.
\end{equation}

\noindent The diffusion matrix takes the neutral Moran form
\begin{equation}
B_{ij}=
\begin{cases}
\dfrac{2}{N^2}x_i(1-x_i), & i=j,\\[6pt]
-\dfrac{2}{N^2}x_ix_j, & i\neq j,
\end{cases}
\qquad i,j=1,2.
\end{equation}
Next, we apply the general perturbative approach derived in Sec.~\ref{sec:general-perturbative} to obtain the fixation probability of allele $1$, $\phi_1(x_1,x_2)$. To this end, we first write the decomposition of the Fokker-Planck backward operator for this particular choice of fitness functions,
\begin{eqnarray}\label{eq:L_M3}
\mathcal{L}^{(0)} &=&\frac{1}{2}\Big(
B_{11}\,\partial_{x_1}^2
+2B_{12}\,\partial_{x_1}\partial_{x_2}
+B_{22}\,\partial_{x_2}^2
\Big), \nonumber \\
\mathcal{L}^{(s)}
&=&
A_1\,\partial_{x_1}+A_2\,\partial_{x_2}.
\end{eqnarray}
Next, we replace the fitness functions and expressions for the coefficients $B_{k\ell}$ into the general expression for $\phi_i^{(s)}$ in Eq.\,\eqref{eq:PDE_phi1_general}. This substitution yields an equation for the first-order perturbative correction to the neutral fixation probability,
\begin{equation}\label{eqconstant}
\begin{aligned}
0
&= Nx_1\,(s_1 - s_1 x_1 - s_2 x_2 - s_3 (1 - x_1 - x_2))
\\[3pt]
&\quad+
x_1(1-x_1)\,\partial_{x_1}^2\phi_1^{(s)}
- 2x_1 x_2\,\partial_{x_1}\partial_{x_2}\phi_1^{(s)}
\\[3pt]
&\quad+ x_2(1-x_2)\,\partial_{x_2}^2\phi_1^{(s)},
\end{aligned}
\end{equation}
with $\phi_1^{(s)}=0$ on the absorbing boundaries, as discussed above.

For this particular choice of constant fitness functions, the right-hand side of Eq.~\eqref{eqconstant} is at most quadratic in the allele frequencies. Because the neutral backward Fokker-Planck operator only involves second-order derivatives with coefficients that are at most quadratic in $x_i$, it is sufficient to look for a first-order perturbative correction that is also at most quadratic. The polynomial ansatz can thus be written as
\begin{equation}\label{eq:phi1s_const_ansatz}
\phi_1^{(s)}(x_1,x_2)
=
A\,x_1+B\,x_1^2+C\,x_1x_2,
\end{equation}
where the absence of pure $x_2$ terms follows from the absorbing condition along $x_1=0$. Note also that we have denoted the coefficients by $A$, $B$, and $C$ instead of using the general notation introduced in Section~\ref{sec:general-perturbative} for clarity.

To obtain the coefficients $A$, $B$, and $C$, we compute the derivatives of this polynomial ansatz with respect to each allele frequency and substitute them into Eq.\,\eqref{eqconstant}. This yields the polynomial identity
\begin{equation}\label{eq:const_identity}
\begin{aligned}
0
&= Nx_1\,(s_1 - s_1 x_1 - s_2 x_2 - s_3 (1 - x_1 - x_2))
\\[3pt]
&\quad+
2B\,x_1(1-x_1)-2C\,x_1x_2.
\end{aligned}
\end{equation}
Imposing that each term in the multivariate polynomial on the right-hand side of Eq.\,\eqref{eq:const_identity} vanishes, we obtain
\begin{equation}
B=\frac{N}{2}(s_3-s_1),
\qquad
C=\frac{N}{2}(s_3-s_2).
\end{equation}
The remaining coefficient $A$ follows from the boundary condition $\phi_1^{(s)}(1,0)=0$,
\begin{equation}
A=\frac{N}{2}(s_1-s_3).
\end{equation}

Substituting these coefficients into Eq.\,\eqref{eq:phi1s_const_ansatz} and combining the resulting expression for $\phi_1^{(s)}$ with the neutral fixation probability, we obtain
\begin{equation}\label{eq:phi1_const_final}
\phi_1(x_1,x_2)
=
x_1+\frac{Nx_1}{2}\big(s_1-\bar s\big).
\end{equation}
This solution is valid to first order in $s_i=\mathcal{O}(N^{-1})$.

Although Eq.\,\eqref{eq:phi1_const_final} was obtained explicitly for three alleles, its form remains unchanged for an arbitrary number $M$ of competing alleles. To show this, we consider
\begin{equation}
f_j=1+s_j,
\qquad
\bar s=\sum_{j=1}^{M}x_js_j,
\end{equation}
with $x_M=1-\sum_{k=1}^{M-1}x_k$. Eliminating $x_M$, the mean selection coefficient becomes
\begin{equation}
\bar s
=
s_M+\sum_{k=1}^{M-1}x_k(s_k-s_M).
\label{eq:sbar_general_const}
\end{equation}
For any focal allele $i=1,\ldots,M-1$, the general perturbative equation \eqref{eq:PDE_phi1_general} then reduces to
\begin{equation}
\mathcal{L}^{(0)}\phi_i^{(s)}
=
-\frac{x_i}{N}\big(s_i-\bar s\big).
\label{eq:general_const_PDE}
\end{equation}

Motivated by the three-allele result, we propose
\begin{equation}
\phi_i^{(s)}
=
\frac{Nx_i}{2}\big(s_i-\bar s\big).
\label{eq:general_const_correction}
\end{equation}
We can verify this expression directly. Using Eq.\,\eqref{eq:sbar_general_const}, the polynomial inside the correction can be written as
\begin{equation}
x_i(s_i-\bar s)
=
x_i(s_i-s_M)
-
\sum_{k=1}^{M-1}
x_ix_k(s_k-s_M).
\label{eq:general_const_expansion}
\end{equation}
The first term on the right-hand side is linear in the allele frequencies and is therefore annihilated by $\mathcal{L}^{(0)}$, which contains only second derivatives. For the quadratic terms, the neutral operator gives
\begin{equation}
\mathcal{L}^{(0)}(x_ix_k)
=
\begin{cases}
\dfrac{2}{N^2}x_i(1-x_i), & k=i,\\[6pt]
-\dfrac{2}{N^2}x_ix_k, & k\neq i.
\end{cases}
\label{eq:neutral_quadratic_terms}
\end{equation}
Applying these two results to Eq.\,\eqref{eq:general_const_expansion}, we obtain
\begin{align}
\mathcal{L}^{(0)}
\left[x_i(s_i-\bar s)\right]
&=
-\frac{2x_i}{N^2}(s_i-s_M)(1-x_i)
\nonumber\\
&\quad+
\frac{2x_i}{N^2}
\sum_{\substack{k=1\\k\neq i}}^{M-1}
x_k(s_k-s_M)
\nonumber\\
&=
-\frac{2x_i}{N^2}\big(s_i-\bar s\big).
\label{eq:general_const_operator_result}
\end{align}
Multiplying this result by $N/2$ gives exactly the right-hand side of Eq.\,\eqref{eq:general_const_PDE}. Therefore, Eq.\,\eqref{eq:general_const_correction} solves the perturbative equation for any number of alleles. It also satisfies the required boundary conditions: the correction vanishes when $x_i=0$ and when allele $i$ is fixed, since $\bar s=s_i$ at $x_i=1$.

Combining the perturbative correction with the neutral solution gives
\begin{equation}\label{eq:phi_const_general}
\phi_i(\mathbf{x})
=
x_i+\frac{Nx_i}{2}\big(s_i-\bar s\big),
\qquad
i=1,\ldots,M.
\end{equation}
The general solution also gives a simple condition for when selection favors a focal allele. Because an allele's neutral fixation probability is $x_i$, $\phi_i(\mathbf{x})>x_i$ if and only if $s_i>\bar s$. Using the fact that the population mean fitness is $\bar f=1+\bar s$ and $f_i=1+s_i$, imposing $s_i>\bar s$ is equivalent to imposing $f_i>\bar f$. Thus, for any number of competing alleles, an allele has a fixation probability greater than its neutral value whenever its fitness is greater than the mean fitness of the initial population. Within the weak-selection regime, this criterion does not depend explicitly on the population size or on the number of competing alleles. This simple rule illustrates how the analytical solution turns a high-dimensional stochastic problem into a direct biological statement.

\begin{figure}[h]
\centering
\includegraphics[width=\columnwidth]{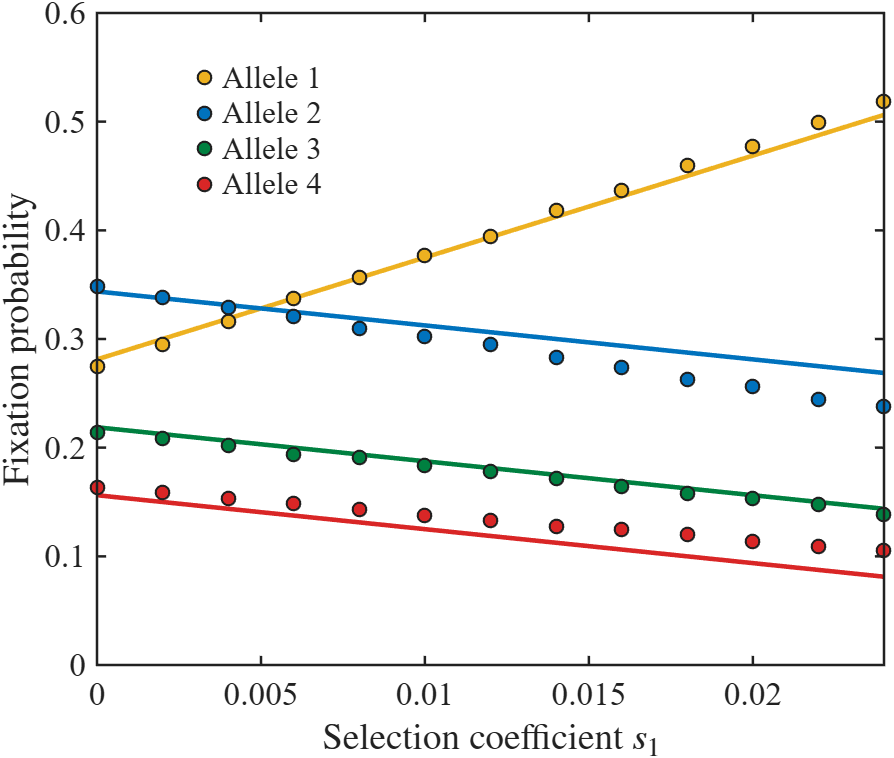}
\caption{Fixation probabilities in a four-allele Moran process with constant fitness. Symbols correspond to Monte Carlo simulations of the discrete process ($10^5$ realizations per value of $s_1$) and lines to the general weak-selection solution in Eq.~\eqref{eq:phi_const_general}. The agreement remains excellent even for $s_1$ above the formal weak-selection scale $\mathcal{O}(1/N)$. Parameters: $N=100$, $s_2=0.005$, $s_3=-0.005$, $s_4=-0.010$, and $x_1=x_2=x_3=x_4=1/4$..}
\label{fig:constfitness}
\end{figure}

The general result in Eq.\,\eqref{eq:phi_const_general} is compared with Monte Carlo simulations of a four-allele discrete Moran process in Fig.~\ref{fig:constfitness}. The analytical solution accurately describes the fixation probabilities of all four alleles as the fitness of allele~1 is varied, providing an explicit validation of the method beyond the three-allele case.

\subsection{Frequency-dependent fitness: coordination game}

We next analyze an example with frequency-dependent fitness, typical of evolutionary game theory. We will focus on a three-allele coordination game and obtain the fixation probability for allele $1$. Our approach is, however, straightforwardly generalizable to any multivariate polynomial fitness function, as shown in Section \ref{sec:general-perturbative}.

In the coordination game, each species' fitness increases linearly with its frequency in the population. Hence, $\pi_i(\mathbf{x})=p_ix_i$, and the fitness functions can be written as
\begin{equation}
f_1 = 1 + s_1 x_1,\qquad
f_2 = 1 + s_2 x_2,\qquad
f_3 = 1 + s_3 x_3,
\end{equation}
where $s_i\equiv s p_i<\mathcal{O}(N^{-1})$.

Coordination games arise in natural scenarios in which a phenotype becomes more advantageous as more individuals adopt the same strategy, and are common in microbial systems. For example, invertase production in \textit{Saccharomyces cerevisiae} confers fitness benefits only when producers exceed a critical frequency, leading to bistable dynamics characteristic of coordination \cite{Gore2009}. Similar frequency-dependent benefits occur in \textit{Pseudomonas aeruginosa}, where quorum-sensing-controlled elastase production enhances growth above a signaler density threshold \cite{Darch2012}. Finally, coordination-like dynamics can also appear in animal systems, such as fish schools exhibiting collective escape responses, where the fitness benefit of reacting early depends strongly on the number of neighbors adopting the same strategy \cite{Handegard2010}.

\begin{figure}[h]
\centering
\includegraphics[width=8cm]{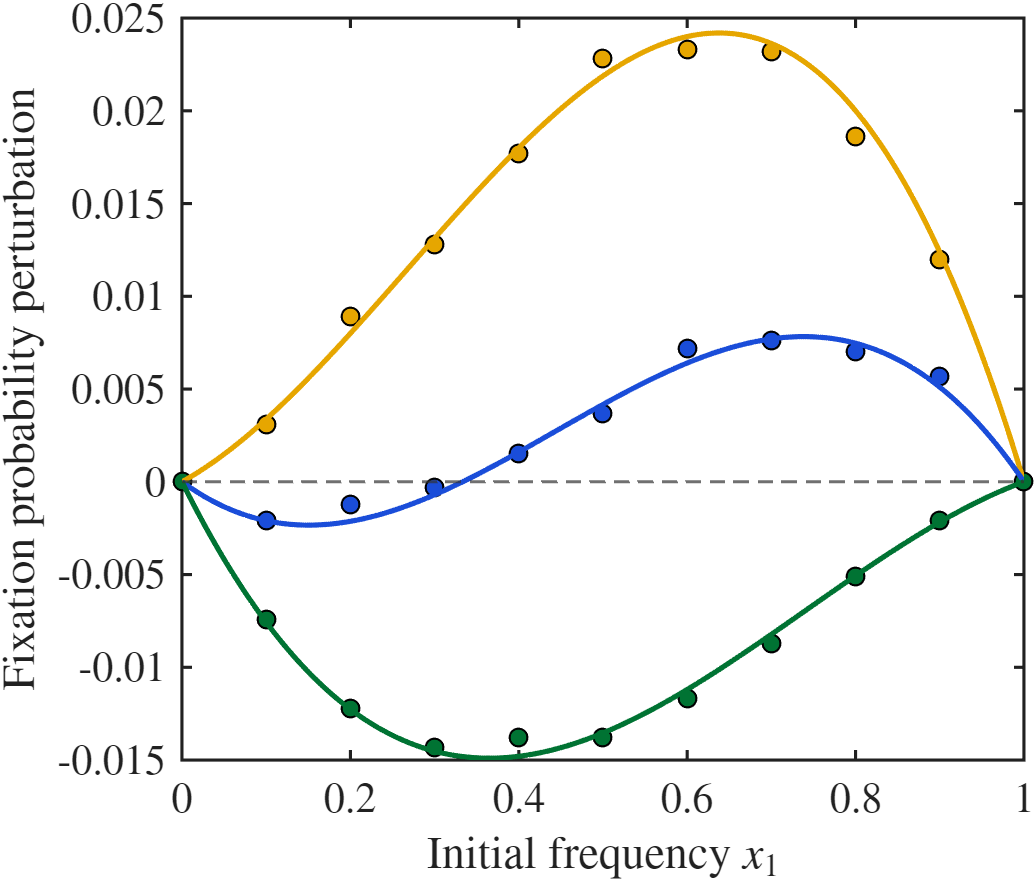}
\caption{Deviation of allele-1 fixation probability from neutrality, $\phi_1^{(s)}=\phi_1-x_1$, in a three-allele Moran process with coordination-game fitness. Symbols show Monte Carlo simulations ($10^6$ realizations), and solid curves show the weak-selection analytical prediction \eqref{eq:phi1_coord_final}. Initial conditions are $(x_1,x_2,x_3)=(x_1,\tfrac{1-x_1}{2},\tfrac{1-x_1}{2})$, with $x_1\in[0,1]$ sampled in steps of $0.1$. Selection coefficients are $(s_1,s_2,s_3)=10^{-3}(6,4,2)$ (yellow), $10^{-3}(4,6,2)$ (blue), and $10^{-3}(2,6,4)$ (green).}
\label{fig:coord_fig}
\end{figure}

Following the general steps introduced in Section \ref{sec:general-perturbative} and replicated for the constant-fitness case in \ref{subsec:const}, we first obtain a partial differential equation for the first-order perturbative contribution to the fixation probability $\phi_1^{(s)}$,
\begin{equation}\label{eq:coord_PDE}
\begin{aligned}
0
&= Nx_1\,(s_1x_1 - s_1 x_1^2 - s_2 x_2^2 - s_3 (1 - x_1 - x_2)^2)
\\[3pt]
&\quad+
x_1(1-x_1)\,\partial_{x_1}^2\phi_1^{(s)}
- 2x_1 x_2\,\partial_{x_1}\partial_{x_2}\phi_1^{(s)}
\\[3pt]
&\quad+ x_2(1-x_2)\,\partial_{x_2}^2\phi_1^{(s)}.
\end{aligned}
\end{equation}
with boundary conditions $\phi_1^{(s)}(0,x_2)=0$ and $\phi_1^{(s)}(1,0)=0$. This equation suggests a multivariate polynomial ansatz of the form
\begin{equation}\label{eq:coord_ansatz}
\begin{aligned}
\phi_1^{(s)} = N\big(&Ax_1 + Bx_1^2 + Cx_1^3 \\
&+ Dx_1x_2 + E x_1^2 x_2 + F x_1x_2^2\big),
\end{aligned}
\end{equation}
where, again, the absence of pure $x_2$ terms follows from the absorbing condition along $x_1=0$. Computing the relevant derivatives of this ansatz with respect to allele frequencies, and substituting them in Eq.\,\eqref{eq:coord_PDE}, we obtain
\begin{equation}\label{eq:coord_identity}
\begin{aligned}
0
&=
(2B - s_3)x_1 + (6C - 2B + s_1 + 2s_3)x_1^2
\\[3pt]
&\quad-
(6C + s_1 + s_3)x_1^3 + (2E + 2F - 2D + 2s_3)x_1x_2
\\[3pt]
&\quad-
(6E + 2s_3) x_1^2 x_2 - (6F+ s_2 + s_3) x_1x_2^2.
\end{aligned}
\end{equation}
Finally, by making each order in the multivariate polynomial vanish, we obtain the coefficients $B$, $C$, $D$, $E$, $F$, while $A$ follows from imposing the boundary condition $\phi_1^{(s)}(1, 0) = 0$. These steps lead to a final form for the allele $1$ fixation probability
\begin{equation}\label{eq:phi1_coord_final}
\begin{aligned}
\phi_1
&=
x_1 + \frac{Nx_1}{6} \left[ s_1 (1-x_1^2) - s_2(x_2 + x_2^2)\right]
\\[3pt]
&\quad-
\frac{Nx_1}{6}s_3\big(2 - 3x_1 + x_1^2 - 3x_2 + 2x_1 x_2 + x_2^2 \big),
\end{aligned}
\end{equation}
which is, again, in excellent agreement with Monte Carlo simulations in the weak-selection regime (Fig.~\ref{fig:coord_fig}).

\subsection{Frequency-dependent fitness: clonal interference between mutualistic alleles}

A different frequency-dependent scenario arises when two alleles are each selectively inferior to a third in pairwise competition, but can jointly overcome this disadvantage through mutualistic fitness feedback. In such situations, the fixation outcomes depend not only on the intrinsic selective advantage of a new mutation, but also on whether another beneficial lineage is rising at the same time, a phenomenon known as clonal interference \cite{GerrishLenski1998,DesaiFisher2007}. In microbial evolution, mutations that would fix in isolation may be blocked by the emergence of a competing lineage, while otherwise deleterious mutations can be rescued by the appearance of another type that interacts positively with it. Thus, mutualistic feedback can change the evolutionary trajectories and allow cooperative alleles to challenge an otherwise dominant competitor.

For a realization of this mutualistic setup, we consider a positive fitness feedback between alleles~1 and 2 and constant selection for allele~3. In this case, the fitness functions can be written as
\begin{equation}\label{eq:fitness-clonal}
f_1 = 1 + s_2 x_2,\qquad
f_2 = 1 + s_1 x_1,\qquad
f_3 = 1 + s_3,
\end{equation}
with $s_i=sp_i$.
\begin{figure*}
 \includegraphics[width=\textwidth]{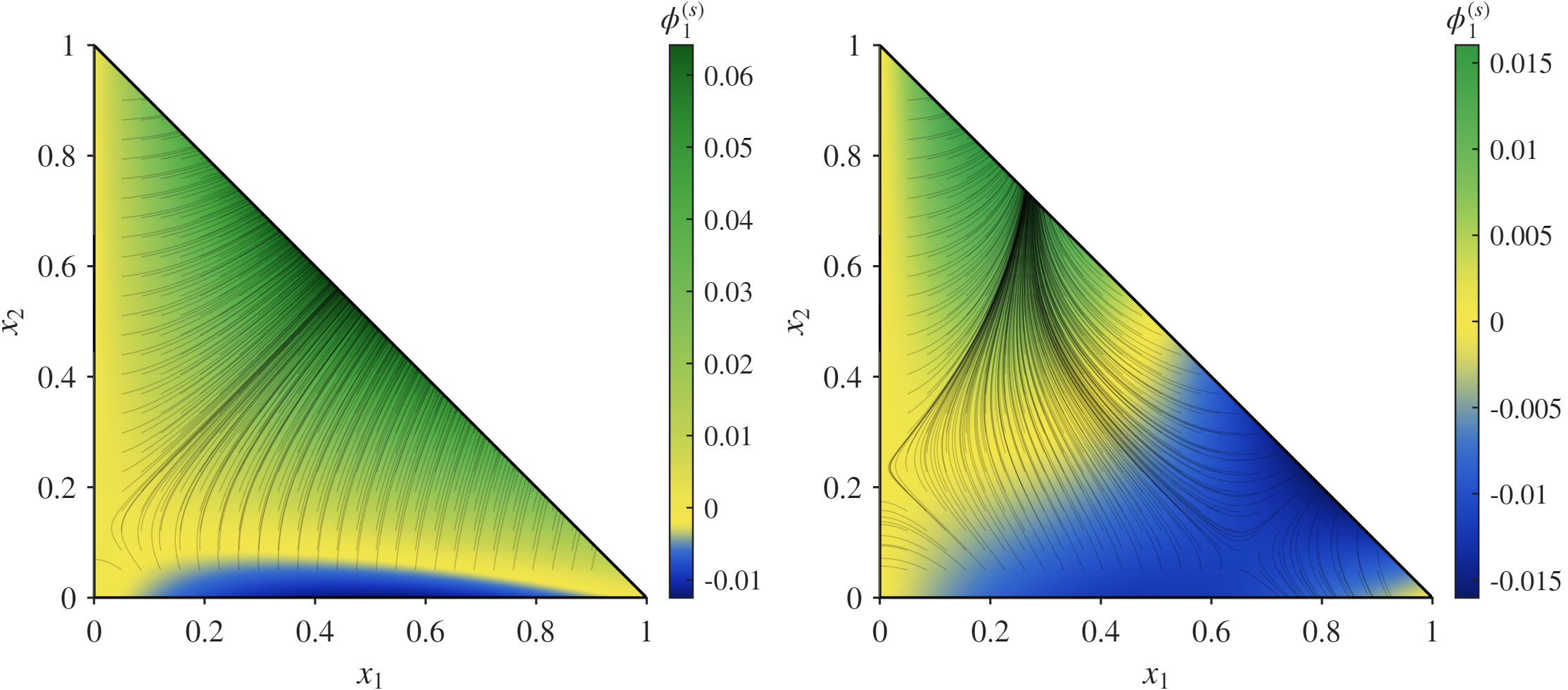}
\caption{Fixation probability perturbation $\phi_1^{(s)}=\phi_1 - x_1$ for allele~1 in the three-allele Moran process under frequency-dependent selection. The color map represents the analytical solution over the simplex, while streamlines indicate the direction of maximal increase of the perturbation. (a) In the one-sided interaction regime, $s_1=0$, gradient flows largely follow directions in which both $x_1$ and $x_2$ increase. (b) With mutualistic feedback, $s_1=0.01$, the geometry of the perturbation changes qualitatively and the gradient flow bends into regions where the fixation advantage of allele~1 increases as $x_1$ decreases, reflecting effective competition between alleles~1 and~2 once allele~3 becomes rare. Other parameter values: $s_2=0.01$, $s_3=0.001$ and $N=100$ in both panels.}
\label{fig:mutualism}
\end{figure*}

For the fitness functions in Eq.\,\eqref{eq:fitness-clonal}, the equation for $\phi_1^{(s)}$ becomes
\begin{equation}\label{eq:mutual_PDE}
\begin{aligned}
0
&= Nx_1\,[(1-x_1)(s_2x_2 - s_3) + x_2(s_3 - s_1x_1)]
\\[3pt]
&\quad+
x_1(1-x_1)\,\partial_{x_1}^2\phi_1^{(s)}
- 2x_1 x_2\,\partial_{x_1}\partial_{x_2}\phi_1^{(s)}
\\[3pt]
&\quad+ x_2(1-x_2)\,\partial_{x_2}^2\phi_1^{(s)}.
\end{aligned}
\end{equation}
with boundary conditions $\phi_1^{(s)}(0,x_2)=0$ and $\phi_1^{(s)}(1,0)=0$. The multivariate polynomial ansatz has the same structure we used for the coordination game, \eqref{eq:coord_ansatz}. Computing its derivatives and plugging them in Eq.\,\eqref{eq:mutual_PDE}, we obtain the identity
\begin{equation}\label{eq:mutual_identity}
\begin{aligned}
0
&=
(2B - s_3)x_1 + (6C - 2B + s_3)x_1^2
\\[3pt]
&\quad-
6Cx_1^3 + (2E + 2F - 2D + s_2 + s_3)x_1x_2
\\[3pt]
&\quad-
(6E +s_1 + s_2) x_1^2 x_2 - 6F x_1x_2^2,
\end{aligned}
\end{equation}
 from which we can obtain the coefficients $B$, $C$, $D$, $E$, $F$ by making each order in the multivariate polynomial vanish. As in previous examples, $A$ follows from the boundary condition $\phi_1^{(s)}(1, 0) = 0$. These steps lead to a final expression for the fixation probability

\begin{align}\label{eq:phi1_mutual_final}
\phi_1
&=
x_1 + \frac{Nx_1}{6} \left[-s_1 (x_2 + x_1x_2) + s_2(2x_2 - x_1x_2)\right] \nonumber \\
&\quad+
\frac{Nx_1}{2}s_3(x_1 + x_2 -1).
\end{align}

The perturbative correction in Eq.\,\eqref{eq:phi1_mutual_final}, $\phi_1-x_1$, quantifies the deviation of the fixation probability from neutrality caused by the mutualistic interaction between alleles $1$ and $2$. When the mutualism is asymmetric (i.e., $s_1=0$), the deviation from neutrality increases monotonically with both $x_1$ and $x_2$ (Fig.~\ref{fig:mutualism}A). In this case, a higher frequency of allele~2 enhances the fitness of allele~1, and this effect is amplified when allele~1 is already at a higher frequency, leading to progressively larger positive deviations from neutrality.

On the other hand, when $s_1 > 0$ and alleles~1 and~2 mutually enhance each other's fitness, the structure of the fixation probability correction changes qualitatively
(Fig.~\ref{fig:mutualism}B). The largest fixation benefits for allele~1 no longer occur just at large $x_1$. They are instead observed along directions in the simplex where alleles~1 and~2 coexist at intermediate frequencies. This shift occurs because increasing the frequency of allele~1 also strengthens allele~2, which in turn competes with allele~1 for fixation. As a result, the selective flow bends away from purely increasing $x_1$ directions. This interplay generates a nontrivial geometric structure in the simplex and illustrates how mutualistic clonal interference reshapes fixation probabilities.

The dependence of $\phi_1^{(s)}$ on the initial composition also suggests how the mutualistic feedback could be detected in a competition experiment that measures changes in the fixation probability of one allele due to the presence of its partner. A simple setup to measure this effect would require two populations started from the same composition $(x_1,x_2,x_3)$. In one of them, alleles 1 and 2 interact through the mutualistic feedback of Eq.\,\eqref{eq:fitness-clonal}. In the other, allele 2 is replaced by a strain that does not interact with allele 1, so $f_1=f_2=1$ and $f_3=1+s_3$. Allele 1's fixation probability in this second, control, population is given by the constant-fitness solution in Eq.\,\eqref{eq:phi_const_general},
\begin{equation}\label{eq:phi1_ctrl}
\phi_1^{\rm ctrl}=x_1-\frac{Nx_1}{2}\,s_3\,(1-x_1-x_2),
\end{equation}
which we can subtract from Eq.\,\eqref{eq:phi1_mutual_final} to obtain
\begin{equation}\label{eq:delta_phi1}
\Delta\phi_1=\frac{Nx_1x_2}{6}\big[s_2(2-x_1)-s_1(1+x_1)\big].
\end{equation}
The neutral value $x_1$ and every term in $s_3$ cancel. For symmetric feedback, $s_1=s_2=s$, the difference reduces to $Nsx_1x_2(1-2x_1)/6$ and changes sign at $x_1=1/2$. The partner raises the fixation probability of allele~1 when it starts below half of the population and lowers it above, which is the mutualistic contribution to $\phi_1^{(s)}$ in Fig.~\ref{fig:mutualism}b.

\section{Discussion}

We introduced a perturbative method to obtain fixation probabilities in general multi-allele Moran processes under weak selection. The method exploits the general structure of the backward Fokker-Planck operator in this limit to obtain fixation probabilities following a systematic expansion of these quantities around their neutral solution. When fitness functions are multivariate polynomials in allele frequencies, this expansion reduces the high-dimensional boundary-value problem that defines fixation probabilities to a finite linear algebraic system that can be solved for an arbitrary number of alleles \cite{Gardiner2009}.

Additionally, because the backward Fokker–Planck operator in the weak-selection regime always has a multivariate polynomial structure, it should be straightforward to extend this approach to evolutionary dynamics in which fitness functions are multivariate polynomials in allele frequencies. These include classical linear payoff matrices from evolutionary game theory \cite{Nowak2006} and nonlinear interactions arising in ecological and genetic models, such as higher-order interactions or context-dependent fitness landscapes \cite{Grilli2017}. Smooth non-polynomial fitness functions may still be treated by approximating them through a truncated multivariate Taylor expansion over the relevant region of the allele-frequency simplex. When this truncation provides a good approximation to the original fitness functions, the resulting polynomial can be used directly in our framework. This introduces an additional truncation error and may require a larger number of coefficient-matching equations as higher-order terms are retained.

Although the number of coefficients to be obtained increases with the number of alleles and the degree of the polynomial fitness function, it does not depend on the population size, and the underlying perturbative construction of the solution remains unchanged. These two dependencies set a key difference from previous approaches. The exact discrete formulation holds for any selection intensity, but it has one unknown per state, $(N-1)(N-2)/2$ of them for three alleles \cite{FerreiraNeves2018}, while ours carries only the $(d+2)^{M-1}$ coefficients of the ansatz. Both numbers of unknowns scale exponentially with the number of alleles, but the base is $d+2$ instead of $N$. The perturbative solution for the Wright-Fisher model requires expected products of allele frequencies under neutrality, with more frequencies entering these products as the degree of the fitness function increases \cite{LessardLahaie2009}. Our coefficient-matching method does not change with the degree. Our method is therefore limited to weak selection, and within that regime it gives closed expressions for the fixation probability over the whole simplex.

Beyond providing explicit expressions, our framework enables a geometric interpretation of fixation in multi-allele systems. In the diffusion approximation, fixation probabilities define scalar fields over the $(M-1)$-dimensional simplex, whose gradients quantify how weak selection biases stochastic evolutionary outcomes. Perturbative corrections thus not only determine the magnitude of deviations from neutrality, but also the structure of selection-induced directional biases across the allele-frequency space. In particular, the nontrivial dependence of these gradients on population composition, as observed in our examples, shows how selection reshapes fixation tendencies in the allele-frequency simplex and induces nontrivial directional biases in stochastic evolutionary dynamics.

This geometric interpretation of fixation probabilities provides a natural way to understand multi-allele effects in the underlying evolutionary dynamics. Across various applications to three-allele Moran processes, we showed that fixation dynamics in these high-dimensional systems cannot, in general, be reduced to a pairwise scenario. Even in the simplest case with constant fitness, the fixation probability of a focal allele depends on the full population composition through the mean fitness, leading to situations in which an allele can benefit from the presence of a third, less-fit competitor. When fitness functions are frequency-dependent, interactions among multiple strains have a richer effect on fixation probabilities \cite{FerreiraNeves2018, LessardLahaie2009}. For example, in the mutualistic clonal interference scenario, the interplay between cooperation and competition generates fixation patterns that are qualitatively different from any two-allele analog. In the mutualistic model, this difference between two-allele and three-allele dynamics can be measured experimentally. Two assays started from the same composition, one with the mutualistic partner and one with a non-interacting strain replacing one of the mutualistic partners, share the same neutral fixation probability, so the difference between them isolates the feedback. The effect is of order $Ns$, so resolving it requires many independent populations per condition.

In addition to providing an analytical understanding of fixation outcomes in multi-allele Moran processes, our results open several directions for future research. The most immediate extension is to relax the weak-selection assumption. Using asymptotic approaches, such as WKB methods for solving the master equation \cite{AssafMeerson2010}, or accounting for higher-order corrections in our perturbative expansion of the fixation probability could help close the gap between analytically tractable weak-selection dynamics and stronger-selection regimes. A second important direction is the extension of this formalism to structured populations \cite{Lieberman2005}, in which the diffusion approximation generally yields effective drift and diffusion terms that depend on network topology or spatial correlations \cite{Allen2017}. Our perturbative approach could be adapted to these spatially structured systems, provided that the weak-selection drift retains a multivariate polynomial structure that allows the order-matching solving strategy to be applied.

In summary, we developed a perturbative framework to compute fixation probabilities in multi-allele Moran processes and applied it to three biological scenarios drawn from evolutionary game theory, microbial ecology \cite{Czaran2002}, and animal behavior \cite{HofbauerSigmund1998}. More broadly, this method contributes to current efforts to develop analytical tools for understanding stochastic evolutionary models \cite{Traulsen2007}, where high-dimensional problems have been studied mainly through numerical simulations \cite{AllenTarnita2014}. Our results thus extend analytical access to fixation probabilities beyond pairwise interactions, enabling a deeper mathematical understanding of multi-strategy stochastic evolutionary dynamics.

\section*{Acknowledgements}
This work was partially funded by the Center of Advanced Systems Understanding (CASUS), which is financed by Germany’s Federal Ministry of Research, Technology, and Space (BMFTR) and by the Saxon Ministry for Science, Culture, and Tourism (SMWK) with tax funds on the basis of the budget approved by the Saxon State Parliament. IB acknowledges financial support from the Conselho Nacional de Desenvolvimento Científico e Tecnológico (CNPq) through a postdoctoral fellowship with grant number 401209/2024-5. RMG received partial support from the São Paulo Research Foundation (FAPESP) through grant ICTP-SAIFR 2021/14335-0.

\onecolumngrid

\begin{appendix}

\section{Number of equations in the polynomial solution}
\label{sec:appPolynomialScaling}

Here, we quantify the number of equations that must be solved in the polynomial construction introduced in Sec.~\ref{sec:general-perturbative}. Following the convention used in the main text, we assume that the fitness deviations $\pi_i(\mathbf{x})$ are multivariate polynomials in which each of the $M-1$ independent allele frequencies appears with degree at most $d$.

Since the mean fitness deviation is $\bar{\pi}(\mathbf{x})=\sum_jx_j\pi_j(\mathbf{x})$, each variable can appear in $\bar{\pi}(\mathbf{x})$ with degree at most $d+1$. The same is therefore true for $\pi_i(\mathbf{x})-\bar{\pi}(\mathbf{x})$. The right-hand side of Eq.\,\eqref{eq:PDE_phi1_general} contains an additional factor $x_i$, which also ensures that the selective contribution vanishes when the focal allele is absent.

The neutral backward operator contains second derivatives multiplied by coefficients that are at most quadratic in the allele frequencies. Consequently, it maps polynomials into polynomials without increasing the maximum degree required in each variable. We can therefore write
\begin{equation}
\phi_i^{(s)}(\mathbf{x})
=
N x_i
\sum_{\beta_1=0}^{d+1}
\cdots
\sum_{\beta_{M-1}=0}^{d+1}
C_{i,\beta_1,\ldots,\beta_{M-1}}
x_1^{\beta_1}\cdots x_{M-1}^{\beta_{M-1}},
\label{eq:app_polynomial_ansatz}
\end{equation}
where the coefficients $C_{i,\beta_1,\ldots,\beta_{M-1}}$ are of first order in the selection strength.

Each exponent $\beta_k$ can independently take the values $0,1,\ldots,d+1$, giving $d+2$ possibilities for each of the $M-1$ independent allele frequencies. The maximum number of unknown coefficients for one focal allele is therefore
\begin{equation}
K(M,d)
=
(d+2)^{M-1}.
\label{eq:number_coefficients}
\end{equation}
This is an upper bound because, for a particular fitness function, some monomials may not be generated and their coefficients are simply zero. For example, in the three-allele constant-fitness case, $M=3$ and $d=0$, and the general construction allows the four terms $x_1$, $x_1^2$, $x_1x_2$, and $x_1^2x_2$. The last term is not generated by the perturbative equation, leaving precisely the three-term ansatz used in the main text.

Substituting Eq.\,\eqref{eq:app_polynomial_ansatz} into Eq.\,\eqref{eq:PDE_phi1_general} and matching equal monomials produces a finite linear system for the unknown coefficients. The absorbing boundary conditions fix the remaining degrees of freedom. Since only $M-1$ fixation probabilities are independent, the complete problem contains at most
\begin{equation}
K_{\mathrm{tot}}(M,d)
=
(M-1)(d+2)^{M-1}
\end{equation}
unknown coefficients, separated into $M-1$ independent systems.

For a fixed polynomial degree, the number of coefficients grows as $(d+2)^{M-1}$ with the number of alleles, whereas, for fixed $M$, it grows polynomially with $d$. Importantly, this number does not depend on the population size $N$. Once the coefficients are determined, the resulting expression gives the fixation probability for every initial population composition. In contrast, Monte Carlo estimates require many complete stochastic realizations for each initial condition. For low-degree fitness functions and moderate numbers of alleles, solving the polynomial system can therefore be orders of magnitude faster than estimating the complete fixation-probability field through simulations.

\section{Fokker-Planck equation in a multi-allele Moran process}
\label{sec:appKramersMoyal}

We consider a Moran process with $M$ alleles in a population of size $N$. The state of the system is given by the vector $\mathbf{n} = (n_1,\dots,n_{M-1})$ and the abundance of the $M$-th allele follows from the conservation of the total population size, $n_M = N - \sum_{k=1}^{M-1} n_k$. Using the unit vectors introduced in the main text, $\bm{\delta}_i$ with components $(\bm{\delta}_i)_k=\delta_{ki}$ (where $\delta_{ki}$ is Kronecker delta), the master equation for the probability of observing the system in state $\mathbf{n}$ is
\begin{equation}\label{eq:masterapp}
\frac{\partial}{\partial t}P(\mathbf{n},t)
=
\sum_{\substack{i,j=1 \\ i\neq j}}^{M}
T_{i\to j}(\mathbf{n}-\bm{\delta}_i+\bm{\delta}_j)
P(\mathbf{n}-\bm{\delta}_i+\bm{\delta}_j,t)
-
\sum_{\substack{i,j=1 \\ i\neq j}}^{M}
T_{i\to j}(\mathbf{n})P(\mathbf{n},t),
\end{equation}
which can be written in the form of Eq.\,\eqref{eq:master} using the step operators in Eq.\,\eqref{eq:step}.

To obtain the Fokker-Planck approximation to Eq.\,\eqref{eq:masterapp}, we define the allele frequencies,
\begin{equation}
x_k = \frac{n_k}{N}, \qquad k=1,\dots,M-1, \nonumber
\end{equation}
for which conservation of the total population becomes, $x_M = 1 - \sum_{k=1}^{M-1} x_k$. In these new variables, we define a rescaled probability density function of observing the system in a state $\mathbf{x}$
\begin{equation}
\rho(\mathbf{x},t) = N^{M-1}P(\mathbf{n},t). \nonumber
\end{equation}

\noindent Defining the family of vectors $\bm{\Delta}^{(i,j)}$ with components
\begin{equation}
(\bm{\Delta}^{(i,j)})_k = \frac{1}{N}(\delta_{k,j}-\delta_{k,i}),
\qquad k=1,\dots,M-1, \nonumber
\end{equation}
the master equation \eqref{eq:masterapp} for $\rho(\mathbf{x},t)$ becomes
\begin{equation}
\frac{\partial}{\partial t}\rho(\mathbf{x},t)
=
\sum_{\substack{i,j=1 \\ i\neq j}}^{M}
\left[
T_{i\to j}(\mathbf{x}+\bm{\Delta}^{(i,j)})
\rho(\mathbf{x}+\bm{\Delta}^{(i,j)},t)
-
T_{i\to j}(\mathbf{x})\rho(\mathbf{x},t)
\right]. \nonumber
\end{equation}

In the limit of large $N$, the vectors $\bm{\Delta}^{(i,j)}$ represent small increments, and we can Taylor expand the replacement rates $T_{i\to j}(\mathbf{x})$ and the probability density function $\rho(\mathbf{x},t)$ around $\mathbf{x}$. Retaining only terms up to order $\mathcal{O}(N^{-3})$ and introducing implicit summation over $k,\ell=1, \ldots M-1$ to simplify the notation, we obtain
\begin{equation}
\frac{\partial}{\partial t}\rho(\mathbf{x},t)
=
\sum_{\substack{i,j=1 \\ i\neq j}}^{M}
\left[
\Delta_k^{(i,j)}
\partial_k
\big(
T_{i\to j}(\mathbf{x})\rho(\mathbf{x},t)
\big)
+
\frac{1}{2}
\Delta_k^{(i,j)}\Delta_\ell^{(i,j)}
\partial_k\partial_\ell
\big(
T_{i\to j}(\mathbf{x})\rho(\mathbf{x},t)
\big)
\right]. \nonumber
\end{equation}

Finally, we simplify the double sum over $i$ and $j$ using the Kronecker-delta structure of the components of the vectors $\bm{\Delta}^{(i,j)}$. After operating with $\bm{\Delta}^{(i,j)}$ and collecting terms of equal order in $N$, we obtain a contribution of order $\mathcal{O}(N^{-1})$,
\begin{equation}\label{eq:Akapp}
A_k(\mathbf{x})
=
\frac{1}{N}
\sum_{\substack{i=1 \\ i\neq k}}^{M}\big[
T_{k\to i}(\mathbf{x})
-
T_{i\to k}(\mathbf{x})
\big],
\end{equation}

\noindent For the terms of order $\mathcal{O}(N^{-2})$, we obtain
\begin{equation}\label{eq:Bkkapp}
B_{kk}(\mathbf{x})
=
\frac{1}{N^2}
\sum_{\substack{i=1 \\ i\neq k}}^{M}\big[
T_{i\to k}(\mathbf{x})
+
T_{k\to i}(\mathbf{x})\big]
\end{equation}
and
\begin{equation}\label{eq:Bklapp}
B_{k\ell}(\mathbf{x})
=
-\frac{1}{N^2}
\left[
T_{k\to \ell}(\mathbf{x})
+
T_{\ell\to k}(\mathbf{x})
\right],
\qquad
k\neq \ell.
\end{equation}

\noindent These expressions in \eqref{eq:Akapp}--\eqref{eq:Bklapp} correspond to the drift and diffusion elements in Eqs.\,\eqref{eq:Ak}--\eqref{eq:Bkl} in the main text and can be arranged in a multi-dimensional Fokker-Planck equation for an $M$-allele Moran process,
\begin{equation}
\frac{\partial \rho(\mathbf{x},t)}{\partial t} = -\sum_{k=1}^{M-1}\frac{\partial}{\partial x_k}\bigg(A_k(\mathbf{x})\rho(\mathbf{x},t)\bigg) + \frac{1}{2}\sum_{k,\ell=1}^{M-1}\frac{\partial^2}{\partial x_k \partial x_\ell} \bigg(B_{k\ell}(\mathbf{x})\rho(\mathbf{x},t)\bigg). \nonumber
\end{equation}

\section{Weak-selection approximation in a multi-allele Moran process}
\label{sec:appWeakSel}

In this appendix, we provide the calculations to obtain the weak-selection approximation of the Fokker-Planck equation \eqref{eq:FPE}. After time rescaling, the $i \to j$ replacement rates are given by
\begin{equation}
T_{i\to j}(\mathbf{x}) = F_i(\mathbf{x}) D_j(\mathbf{x}), \nonumber
\end{equation}
with fitness-dependent reproduction and constant death,
\begin{equation}
F_i(\mathbf{x}) = \frac{f_i x_i}{\bar f},
\qquad
D_j(\mathbf{x}) = x_j, \nonumber
\end{equation}
where $\bar f = \sum_{\ell=1}^M f_\ell x_\ell$ is the population average fitness. With these definitions, the replacement rates become
\begin{equation}\label{eq:repl-rates}
T_{i\to j}(\mathbf{x}) = \frac{f_i x_i}{\bar f}\, x_j.
\end{equation}

\textbf{Drift vector.}
Replacing Eq.\,\eqref{eq:repl-rates} in Eq.\,\eqref{eq:Ak} for the $k$-th component of the drift vector $A_k(\mathbf{x})$, we obtain
\begin{align}\label{eq:Akrepl}
A_k(\mathbf{x})
&=
\frac{1}{N}
\left[
\frac{f_k x_k}{\bar f}(1-x_k)
-
\frac{x_k}{\bar f}(\bar f - f_k x_k)
\right] \nonumber \\
&=
\frac{1}{N}
\left[
\frac{f_k x_k}{\bar f}
-
x_k
\right]
=
\frac{x_k}{N}
\left(
\frac{f_k}{\bar f} - 1
\right).
\end{align}
To consider the weak-selection limit, we assume that the fitness of the $k$-th allele can be written as
\begin{equation}
f_k(\mathbf{x}) = 1 + s \,\pi_k(\mathbf{x}), \nonumber
\end{equation}
where $\pi_k(\mathbf{x})$ is the fitness of allele $k$ in a population composition $\mathbf{x}$, and $s$ is sufficiently small to ensure that the product $s\,\pi_k(\mathbf{x})$ is of order $\mathcal{O}(N^{-1})$. Under this assumption, defining $\bar \pi(\mathbf{x})= \sum_i x_i \pi_i(\mathbf{x})$, the mean fitness can be written as $\bar f(\mathbf{x}) =1 + s \,\bar \pi(\mathbf{x})$.

Replacing these definitions in the ratio between the fitness of the $k$-th allele and the mean fitness that appears in Eq.\,\eqref{eq:Akrepl} and expanding this ratio to first order in $s$ leads to
\begin{equation}
\frac{f_k(\mathbf{x})}{\bar f(\mathbf{x})}
=
\frac{1+s \pi_k(\mathbf{x})}{1+s\bar \pi(\mathbf{x})}
\simeq
\big[1+s \pi_k(\mathbf{x})\big]\big[1-s\bar \pi(\mathbf{x})\big]
\simeq
1 + s\big[\pi_k(\mathbf{x}) - \bar \pi(\mathbf{x})\big] + \mathcal{O}(s^2). \nonumber
\end{equation}
Substituting this expansion into the drift term finally gives
\begin{equation}
A_k(\mathbf{x})
\simeq
\frac{x_k}{N}s\big[\pi_k(\mathbf{x}) - \bar \pi(\mathbf{x})\big]. \nonumber
\end{equation}

\textbf{Diffusion matrix.}
We next compute the diagonal and off-diagonal elements of the diffusion matrix $B$. We obtain the diagonal terms first. Replacing the expression of the replacement rates \eqref{eq:repl-rates} in \eqref{eq:Bkkapp}, we get
\begin{align}\label{eq:bkkrepl}
B_{kk}(\mathbf{x})
&=
\frac{1}{N^2}
\left[
\frac{f_k x_k}{\bar f}(1-x_k)
+
\frac{x_k}{\bar f}\bigl(\bar f-f_k x_k\bigr)
\right]
\nonumber\\
&=
\frac{1}{N^2}
\left[
x_k+\frac{f_k x_k}{\bar f}-2\frac{f_k x_k^2}{\bar f}
\right]
=
\frac{1}{N^2}
\left[
x_k+\frac{f_k}{\bar f}\,x_k(1-2x_k)
\right].
\end{align}

For the terms with $k\neq\ell$, plugging \eqref{eq:repl-rates} into \eqref{eq:Bklapp} and performing a Taylor expansion up to first order in $s$, leads to
\begin{equation}\label{eq:bklrepl}
B_{k\ell}(\mathbf{x})
=
-\frac{1}{N^2}
\left[
\frac{f_k x_k}{\bar f}\,x_\ell
+
\frac{f_\ell x_\ell}{\bar f}\,x_k
\right]
=
-\frac{x_k x_\ell}{N^2}\,
\frac{f_k+f_\ell}{\bar f},
\qquad
k\neq\ell.
\end{equation}

Finally, to obtain the weak-selection approximation for the diffusion terms, we replace the diffusion coefficients with their neutral Moran-process values. In the weak-selection regime this corresponds to setting $f_i \equiv 1$ and hence $\bar f = 1$, under which Eqs.\,\eqref{eq:bkkrepl} and \eqref{eq:bklrepl} reduce to
\begin{equation}
B_{kk}(\mathbf{x})=\frac{2}{N^2}x_k(1-x_k),
\qquad
B_{k\ell}(\mathbf{x})=-\frac{2}{N^2}x_k x_\ell
\ \ (k\neq\ell). \nonumber
\end{equation}

\end{appendix}

\end{document}